\documentclass{article}
\usepackage{graphicx}
\usepackage{amsmath}
\usepackage{amssymb}
\usepackage{subfigure}

\newcommand{\ddt}{\frac{d}{dt}}
\newcommand{\dt}{\partial_t}
\newcommand{\da}{\partial_a}
\newcommand{\ds}{\partial_s}

\newcommand{\mt}{\tilde{\mu}}

\newcommand{\bx}{{\bf x}}
\newcommand{\bu}{{\bf u}}
\begin{document}
\title{Models of Microbial Dormancy in Biofilms and Planktonic Cultures}
\date{\today}
\author{Bruce P. Ayati\thanks{Department of Mathematics and Program in Applied Mathematical \& Computational Sciences, University of Iowa, Iowa City, IA 52242-1419  ({\tt bruce-ayati@uiowa.edu}).  This material is based upon work supported by the National Science Foundation under Grant No.~DMS-0914514.}   \and Isaac Klapper\thanks{Department of Mathematical Sciences and Center for Biofilm Engineering, Montana State University
({\tt klapper@math.montana.edu}).  This material is based upon work supported by the National Science Foundation under Grant Nos.~DMS-0934696
and DMS-0826975.}}
\maketitle

\begin{abstract}
We present models of dormancy in a planktonic culture and in biofilm, and examine the relative advantage of short dormancy versus long dormancy times in each case.  Simulations and analyses indicate that in planktonic batch cultures and in chemostats, live biomass is maximized by the fastest possible exit from dormancy.    The lower limit of time to reawakening is thus perhaps governed by physiological, biochemical or other constraints within the cells.   In biofilm we see that the slower waker has a defensive advantage over the fast waker due to a larger amount of dormant biomass, without an appreciable difference in total live biomass. Thus it would seem that typical laboratory culture conditions can be unrepresentative of the natural state. We discuss the computational methods developed for this work.
\end{abstract}

%\begin{keywords} 
%biofilm, physiological structure, dormancy, bacteria
%\end{keywords}

%\begin{AMS}
%35B40, 65M25, 92-04, 92-08, 92C37, 92C50
%\end{AMS}
%
%\pagestyle{myheadings}
%\thispagestyle{plain}
%\markboth{B. P. AYATI AND I. KLAPPER}{DORMANCY IN BIOFILM}

\section{Introduction}

Microbial populations, particularly those in biofilms 
(sessile, matrix encased communities, see \cite{KlapperDockerySIREV} for an overview),
can contain cells in varying phenotypic states. 
An important difference between planktonic (free-swimming) and biofilm environments is that
the former is generally well-mixed whereas the latter is unmixed and spatially
heterogeneous as a result. As a result of their self-generated 
spatially variable environmental, biofilms demonstrate spatially
diverse ecology \cite{Zhang1994,StewartFranklin2008}.
Such diversification may be advantageous for 
defense against an uncertain and temporally varying environment. 
For example, though cell states that are more tolerant to antimicrobial 
challenge may be less competitive in the absence of that challenge, their 
presence can improve community survivability against attack.  

Here we consider one such defense mechanism, dormancy (possibly related 
to the phenomenon of persister cells \cite{MMSbiofilm07,persistenceSenescence}) where, in response to an environmental stress, cells 
differentiate into a protected, slow- or non-growing condition 
\cite{Chavez08,Roostalu08}. Bacteria in planktonic states 
have been found to revive faster from dormancy than those in a biofilm state \cite{Chavez08}. 
Thus it would seem that dormancy-regulating parameters
are subject to influence of environmental variation, at least of the sort
found in biofilms. Here we wish to use modeling tools in order to gain
insight into role and regulation of dormancy in spatially mixed systems (batch and
chemostat microbial communities) and unmixed systems (biofilm communities).
Our attention is directed to the relative advantage of short dormancy versus long dormancy times in the cases of batch, chemostat, and biofilm states.

In the process, we also present computational tools designed to 
study dormancy within batch, chemostat, and biofilm population dynamics,
in particular with respect to competitiveness.  These tools are an extension of those discussed in \cite{MMSbiofilm07} for persistence and senescence, primarily in the numerical methods used to solve the more general physiological structure used in this paper, and we expect they will have wider applicability to descriptions of  physiological states in both mixed and unmixed microbial communities .  The physiological structure is represented by a continuous variable.  Compartmentalized
dormancy models have been considered elsewhere \cite{Malik:2006p367, Malik:2008p366}.    

This paper is organized as follows.  We present models of chemostat and batch cultures, and asymptotic analyses of their long-time behavior.   We then derive the biofilm model.  We compute numerical solutions of the model equations for the batch, chemostat and biofilm cultures, and discuss the numerical methods developed for these computations.   We conclude with the implications of our results. 

\section{Models of Dormancy in Chemostat and Batch Cultures}\label{sec:chemobatchmodels}
We introduce $s\in[s_0,s^*]$ to index the dormancy state of individual cells, 
with $s_0$ the value at which cells enter dormancy and $s^*$ the 
value at which cells leave dormancy and become active. Cells progress
through dormancy states with ``speed'' $g(s,c)$, where $c$ is concentration
of relevant chemicals (e.g. substrates or antimicrobials); for example, large
concentrations of substrates and/or small concentrations of
antimicrobials imply larger value of $g$. While dormant, cells do 
not grow and divide; on the other hand, dormant cells are presumed
to be hardier in response to environmental stress. 

Let $t\geq 0$ represent time.  Let $u(t)$ represent the density of
active cells, $v(s,t)$ represent the density of dormant cells, and $c(t)$ be a vector of substrate chemical species concentrations.  Let the operator
 $\partial_y$ denote partial differentiation in the subscript
variable $y$.  The active cell population is modeled by an ordinary differential equation, $t>0$, 
\begin{subequations} \label{planktonic}
\begin{multline}
\ddt u(t) =  \underbrace{b(c)u(t)}_{\text{cell division}} - \underbrace{\mu_{u}(c) u(t)}_{\text{death}}\\+ \underbrace{g(s^*,c) v(s^*,t)}_{\text{exit from dormancy}} 
- \underbrace{h(c)u(t)}_{\text{entrance to dormancy}}
- \underbrace{d_0 u(t)}_{\text{washout}},  \label{sPlanktonic} 
\end{multline}
where $d_0$ is the chemostat dilution rate (roughly, inverse of the time scale
for the chemostat contents to be flushed).
The functions $b$ and $\mu_u$ account for cell division and cell ``death".   We use ``death" as shorthand for all forms of inertness not tied to strategic dormancy on the part of the bacteria.    The function $h \geq 0$ is the dormancy rate of active cells.   Let  $g \geq 0$ (as above) and $\mu_v > 0$  denote the reactivation and death rates, resp., of the dormant cells.   We use a  physiologically structured equation for the dormant cell population,
\begin{align}
\dt v(s,t) + \underbrace{\ds(g(s,c) v(s,t))}_{\text{reactivation kinetics}} &= 
\underbrace{-\mu_v(s,c) v(s,t)}_{\text{death}}-\underbrace{d_0 v(s,t)}_{\text{washout}}
,\label{sv_eqn}\\
g(s_0,c)v(s_0,t) &= \underbrace{h(c) u(t)}_{\text{creation of newly dormant cells}},  \label{sBirth}
\end{align}
for $s_0 < s \leq s^*$. For the substrate chemicals, we have
\begin{equation}
\ddt c(t) = -\underbrace{f(c(t),u(t),v(\cdot,t))}_{\text{chemical reactions}}
+\underbrace{d_0(C_0(t)-c(t))}_{\text{chemostat turnover}},   
 \label{sSubstrate}
\end{equation}
\end{subequations}
where $f$ is the vector of reactions and $C_0(t)$ is the input 
concentration vector from the chemostat tank.  Initial conditions are $u(0) = u_0$, $v(0,s) = v_0(s)$, and $c(0) = c_0$.  The model for a batch culture is obtained from (\ref{planktonic}) by setting $d_0=0$.

\section{Long-Time Behavior}
In this section we examine the long-time behavior of chemostat models for steady and periodic cases.

\subsection{Steady Chemostat}\label{Steady_Chemostat}

We consider the long-time behavior of the steady ($C_0(t)=C_0$) chemostat 
system by studying the time-independent solution of
equations~(\ref{planktonic}). In the steady state $c(t)=c$, so
that we can define a new dormancy coordinate $a\in[0,a^*]$ by
\begin{equation}
a(s) = \int_{s_0}^s\frac{ds'}{g(s',c)},
\end{equation}
with $a^*=a(s^*)$. Setting the time derivative to zero, (\ref{sv_eqn}) 
together with~(\ref{sBirth}) can be solved to obtain
\begin{equation}
v(a) = \frac{h(c)u}{g(a^*,c)}e^{-d_0a}e^{-\int_0^a\mu_v(a',c)da'},
\end{equation}
where $u$ is the steady state value of the active cell density
and $s$ has been replaced by $a$. The first exponential factor
accounts for loss due to washout and the second for loss due
to death. Plugging into~(\ref{sPlanktonic})
and again setting the time-derivative to zero, we obtain
\begin{equation}
0 = u\left[b(c) -\mu_u(c) - d_0 
    - \left(1 - e^{-d_0a^*}e^{-\int_0^{a^*}\mu_v(a,c)da}\right)h(c)\right].
 \label{c_steady}
\end{equation}
If $u\neq0$, then the second factor of~(\ref{c_steady}) provides
an equation for $c$. Assuming that $\mu_u$, $\mu_v$, and $h$ are
all decreasing functions of $c$, we can write that second factor
in the form $b(c)-\bar{\alpha}-\tilde{\alpha}(c)$,  where $\bar{\alpha}$
is a constant and $\tilde{\alpha}(c)$ is a decreasing function of $c$
such that $\tilde{\alpha}(c)\rightarrow0$ as $c\rightarrow\infty$.
Thus, assuming that $b(c)$ is a monotone increasing function of $c$
and that $b(c)>\bar{\alpha}$ for $c$ sufficiently large, then it follows
that there is a unique value of $c$ that solves~(\ref{c_steady}) with
$u\neq0$. However if that value is larger than $C_0$ or if
$b(c)<\bar{\alpha}$ for all $c$, then the only admissible solution
of~(\ref{c_steady}) is $u=0$ (washout).

Finally, given the solution for $c$, equation~(\ref{sSubstrate})
can be solved to obtain the long-time behavior of $u$ by setting the time derivative to zero\footnote{If $u=0$, then~(\ref{sSubstrate}) requires that
$d_0(C_0-c) = f(c,0,0)$; generally $f(c,0,0)=0$ in which case
$c=C_0$.}. In the case that $f$ is monotone increasing in its
arguments, we obtain a unique solution for $u$.

For two (or more) species competing in the same chemostat,
the one that has the steady-state solution with smallest value of substrate $c$
is the only long-time survivor \cite{SmithWalton1995}; 
it excludes the other species by continually reducing substrate until
substrate level is below the others' steady-state requirements.
Thus, from~(\ref{c_steady}) it is apparent that the smaller the
size of $h$ (i.e., the lesser the likelihood of entry to dormancy) and
the smaller the value of $a^*$ (i.e., the shorter the dormancy
period), the more competitive the species. Note that a species which
does not go dormant at all will outcompete an otherwise similar species
which does.

\subsection{Periodic Chemostat}

We consider next the long-time behavior of a periodic chemostat
with input substrate concentration $C_0(\omega t)$ where
$C_0(a+1)=C_0(a)$, and offer asymptotics for two special cases.

\subsubsection{Fast Oscillations}\label{asymptoticsFast}

We suppose that the chemostat oscillation period is short compared
to all other time scales of interest. In this limit, the chemostat
will oscillate many times before the microbial inhabitants can react.
This intuition suggests a multiple time scale expansion with
a slow time $t_1$ and a fast time $t_2$ defined by
\begin{equation}
t_1 =  t, \qquad t_2 =  \epsilon^{-1}t,
\end{equation}
where $\epsilon=\omega^{-1}\ll\tau$ for any inherent
time scale $\tau$ in the system. Note $C_0=C_0(t_2)$. We expand 
\begin{subequations}
\begin{align}
u & =  u_0(t_1,t_2) + \epsilon u_1(t_1,t_2) + \ldots \, ,\\ 
v & =  v_0(s,t_1,t_2) + \epsilon v_1(s,t_1,t_2) + \ldots \, ,\\ 
c & =  c_0(t_1,t_2) + \epsilon c_1(t_1,t_2) + \ldots \, ,
\end{align}
\end{subequations}

We suppose that the solution $(u,v,c)$ approaches periodicity
with period $\omega^{-1}$ for long times, so we look for
a solution independent of slow time $t_1$, i.e.,
\begin{subequations}
\begin{align}
u & =  u_0(t_2) + \epsilon u_1(t_2) + \ldots \, ,\\ 
v & =  v_0(s,t_2) + \epsilon v_1(s,t_2) + \ldots \, ,\\ 
c & =  c_0(t_2) + \epsilon c_1(t_2) + \ldots \, ,
\end{align}
\end{subequations}
In this case, $d/dt=\epsilon^{-1}d/dt_2$. Then to lowest order
($=O(\epsilon^{-1})$), system~(\ref{planktonic}) becomes
\begin{equation}
\frac{\partial u_0}{\partial t_2} =
\frac{\partial v_0}{\partial t_2} =
\frac{\partial c_0}{\partial t_2} =0,
\end{equation}
so that $u_0$, $c_0$ are constants and $v_0=v_0(s)$, i.e.,
\begin{subequations}
\begin{align}
u & =  u_0 + \epsilon u_1(t_2) + \ldots \, ,\\ 
v & =  v_0(s) + \epsilon v_1(s,t_2) + \ldots \, ,\\ 
c & =  c_0 + \epsilon c_1(t_2) + \ldots \, ,
\end{align}
\end{subequations}

At the next order ($=O(\epsilon^0)$), system~(\ref{planktonic}) becomes
\begin{subequations}
\begin{align}
\frac{d}{dt_2} u_1(t_2)  & =  b(c_0)u_0 - \mu_{u}(c_0) u_0 
   + g(s^*,c_0)v_0(s^*) - h(c_0)u_0 - d_0 u_0, \\
\frac{\partial}{\partial t_2} v_1(s,t_2) & =  - \ds(g(s,c_0) v_0(s)) 
-\mu_v(s,c_0) v_0(s)-d_0 v_0(s),\\
\frac{\partial}{\partial t_2} c_1(t_2) & =  
    -f(c_0,u_0,v_0(\cdot))+d_0(C_0(t_2)-c_0),
\end{align}
\end{subequations}
with $g(s_0,c_0)v(s_0) = h(c_0) u_0$. Averaging over a period $\omega^{-1}$,
we obtain
\begin{align}
0  & =  b(c_0)u_0 - \mu_{u}(c_0) u_0 
   + g(s^*,c_0)v_0(s^*) - h(c_0)u_0 - d_0 u_0, \\
0 & =  -\ds(g(s,c_0) v_0(s)) 
-\mu_v(s,c_0) v_0(s)-d_0 v_0(s),\\
0 & =  
    -f(c_0,u_0,v_0(\cdot))+d_0(\bar{C}_0-c_0),
\end{align}
with $g(s_0,c_0)v(s_0) = h(c_0) u_0$, where $\bar{C}_0$ is the average
of $C_0(t_2)$ over one chemostat oscillation period. This 
system, the same as was solved previously in the steady chemostat case
except with $\bar{C}_0$ replacing $C_0$, has essentially the
same solutions for $u_0$, $v_0(s)$,
and $c_0$; the next order terms $u_1(t_2)$, $v_1(s,t_2)$, $c_1(t_2)$ 
add a correction of $O(\epsilon)$. Note thus that the same conclusion
holds: a species without dormancy will outcompete an otherwise
similar species which can go dormant (because the fast oscillating
chemostat acts like a steady chemostat with input substrate $\bar{C}_0$). 

\subsubsection{Slow Oscillations}\label{asymptoticsSlow}

We suppose now that the chemostat oscillation period is long compared
to all other time scales of interest, i.e., that the chemostat
can nearly reach equilibrium before input $C_0(\omega t)$ changes noticeably.
Intuition again suggests a multiple time scale expansion with
a slow time $t_1$ and a fast time $t_2$, in this case defined by
\begin{equation}
t_1 =  \epsilon t, \qquad t_2 =  t,
\end{equation}
where $\epsilon=\omega\ll\tau^{-1}$ for any inherent
time scale $\tau$ in the system. Note that $C_0=C_0(t_1)$. We expand 
\begin{subequations}
\begin{align}
u & =  u_0(t_1,t_2) + \epsilon u_1(t_1,t_2) + \ldots \, ,\\ 
v & =  v_0(s,t_1,t_2) + \epsilon v_1(s,t_1,t_2) + \ldots \, ,\\ 
c & =  c_0(t_1,t_2) + \epsilon c_1(t_1,t_2) + \ldots \, ,
\end{align}
\end{subequations}

We suppose quasi-equilibrium in the sense that $u_0$, $v_0$, and $c_0$ are
independent of fast time $t_2$.
Noting that $d/dt = \epsilon\partial/\partial t_1 + \partial/\partial t_2$,
then at its slowest, $\epsilon^0$ order, system~(\ref{planktonic}) becomes
\begin{subequations}\label{slow_eqns}
\begin{align}
0  & =  b(c_0)u_0 - \mu_{u}(c_0) u_0 
   + g(s^*,c_0)v_0(s^*) - h(c_0)u_0 - d_0 u_0, \\
0 & =  - \ds(g(s,c_0) v_0) 
-\mu_v(s,c_0) v_0-d_0 v_0,\\
0 & =  
    -f(c_0,u_0,v_0(\cdot))+d_0(C_0(t_1)-c_0),
\end{align}
\end{subequations}
with $g(s_0,c_0)v(s_0) = h(c_0) u_0$.
Note that $t_1$ is essentially a parameter, appearing explicitly
only in the input substrate concentration $C_0(t_1)$.
Thus to zeroth order, the quantities $u$, $v$, and $c$ obey
system~(\ref{slow_eqns}), which is the same as the steady
chemostat except with parameterized input substrate.
Hence we again conclude that a species that does not
go dormant will outcompete an otherwise similar one that does.
Note one caveat though: if $C(t_1)$ drops below the minimum
required to sustain a particular population (see 
Section~\ref{Steady_Chemostat}) at any point in its cycle,
then extinction may occur.

\section{A Model of Dormancy in a Biofilm} \label{sec:biofilmmodels}
\begin{figure}[t]
\begin{center}
\includegraphics[height=1.5in]{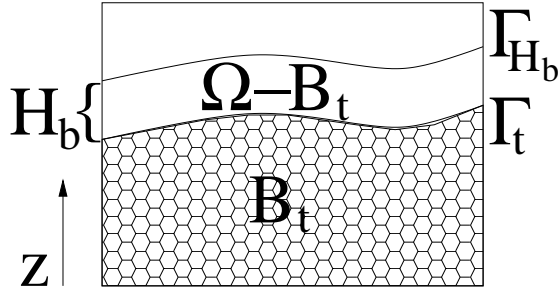}
\caption{Spatial domains for the biofilm model.}
\label{biofilmFig}
\end{center}
\end{figure}

For the biofilm model we remove the chemostat-specific terms and extend the system  (\ref{planktonic}) to include a spatial domain $\Omega$ consisting of stratified subdomains
$B_t$ for biomass and $\Omega \backslash B_t$ for the bulk fluid.
There are two moving interfaces in $\Omega$: $\Gamma_t$ separating
$B_t$ from the rest of $\Omega$, and a bulk-substrate interface
$\Gamma_{H_b}$ that is a fixed height $H_b$ above $\Gamma_t$.  The
biofilm rests on a surface, denoted by a lower boundary, $\Gamma_B$.
The spatial domains are illustrated in Figure \ref{biofilmFig}.   The active and dormant cell populations, and the chemical concentrations, now depend on $\bx \in \Omega$.    Conservation of biomass yields, for $t>0$ and $s_0 < s \leq s^*$,
\begin{subequations} \label{biofilm}
\begin{multline}
\dt u(\bx,t) + \nabla \cdot {\mathbf J_u} = \\  b(c)u(\bx,t) - \mu_u(c) u(\bx,t) + g(s^*,c) v(\bx,s^*,t) - h(c)u(\bx,t),  \label{sBiofilm} 
\end{multline}
\begin{align}
\dt v(\bx,s,t) + \ds(g(s,c) v(\bx,s,t)) + \nabla \cdot {\mathbf J_v} = -\mu_v(s,c) v(\bx,s,t),  &\\
g(s_0,c )v(\bx,s_0,t)  = h(c) u(\bx,t),  \label{sbBirth} &\\
\dt w(\bx,t) + \nabla \cdot {\mathbf J_w} = \mu_u(c) u(\bx,t) + \int_{s_0}^{s^*} \mu_v(s,c) v(\bx,s,t) \ ds, & \\
\dt c(\bx,t) + \nabla \cdot {\mathbf J_c} = f(c(t),u(\bx,t),v(\bx,\cdot,t)),  & \\\end{align}
\end{subequations}
where $\nabla \cdot$ denotes divergence in space, ${\mathbf J_y}$ denotes the flux of subscript variable $y$, and where we assume appropriate initial conditions and boundary condition on the spatial domains.  

Assuming Fick's Law gives ${\bf J_c}= -D \nabla c$ for diffusion constant $D$.  The substrate masses are also subject to
advection, but the velocity is sufficiently slow that we can neglect
the advective contribution to the flux. Likewise, substrate material
diffusion time scales are at least several orders of
magnitude larger than the those at which bacteria grow or advect,
allowing us to make a quasi-steady-state assumption so that
\begin{equation}
-D \nabla^2 c = f.
\end{equation}

Let $\vartheta(\bx,s,t)$ and $\rho(\bx,s,t)$ denote the volume
fraction per dormancy state and density per dormancy state relative to volume fraction, resp., of dormant cells.  We assume incompressibility of
biomass with $\rho(\bx,s,t) \equiv \rho^*$ for positive constant
$\rho^*$.   We also assume, based on the fact that the main constituent
of all cells is water, that active and inert cells have the same incompressibility
properties, and the same densities relative to volume fractions,
$\rho^*$, as dormant cells.  We let $\nu(\bx,t)$ and $\eta(\bx,t)$ denote the volume
fractions of active and dead cells, resp., which are related to the density of active and dead cells by $u = \rho^* \nu$ and $w = \rho^* \eta$.

Assume the biofilm polymer matrix exists in proportion to cell density.  We require the biomass volume fractions to total to one so that
\begin{equation} \label{volumeConstraint}
\nu(\bx,t) + \eta(\bx,t) + \int_{s_0}^{s^*} \vartheta(\bx,s,t) \ ds = 1.
\end{equation}

Assuming that transport of biomass, including dormant cells, is
governed by an advective process, with a volumetric flow $\bu(\bx,t)$
for all classes and ages, gives the fluxes ${\bf J_u} = \rho^* u \bu$, ${\bf J_v} = \rho^* v \bu$ and ${\bf J_w} = \rho^* w \bu$. As in \cite{MMSbiofilm07}, we follow
\cite{multispeciesBiofilm,DockeryKlapperSIAP01} and assume that the
volumetric flow is stress driven according to
\begin{equation}
\bu = -\lambda \nabla p,
  \label{Darcy}
\end{equation}
where $p(t,\bx)$ is the pressure, $\lambda>0$ the Darcy
constant, and $p=0$ in $\Omega \backslash B_t$. Pressure is determined in order
to enforce incompressibility in response to growth 
and hence~(\ref{Darcy}) can be viewed as a balance of growth-induced
stress against friction. Other choices of force balance are possible.

Substituting $u = \rho^* \nu$, ${\bf J_u} = \rho^* u \bu$, $v = \rho^* \vartheta$, ${\bf J_v} = \rho^* v \bu$, $w = \rho^* \eta$,  and ${\bf J_w} = \rho^* w \bu$ into equations (\ref{biofilm}) gives
\begin{subequations} \label{volfracs}
\begin{multline}
\dt \nu(\bx,t) + \nabla \cdot (\bu \nu) = \\ b(c)\nu(\bx,t) -\mu_v(c) \nu(\bx,t) + g(s^*,c) \vartheta(\bx,s^*,t) - h(c)\nu(\bx,t), 
\end{multline}
\begin{align}
\dt \vartheta(\bx,s,t) + \ds(g(s,c) \vartheta(\bx,s,t)) + \nabla \cdot (\bu \vartheta) = -\mu_v(s,c) \vartheta(\bx,s,t),  &   \label{vartheta}\\
g(s_0,c)\vartheta(\bx,s_0,t)  = h(c) \nu(\bx,t), \\
\dt \eta(\bx,t) + \nabla \cdot (\bu \eta) = \mu_u(c) \nu(\bx,t) + \int_{s_0}^{s^*} \mu_v(s,c) \vartheta(\bx,s,t) \ ds, & \\
-D \nabla^2 c = f,& 
\end{align}
\end{subequations}
with appropriate initial conditions and boundary condition on the spatial domains. 

Integrating (\ref{vartheta}) over $s$ gives
\begin{multline}
\underbrace{\dt \left( \int_{s_0}^{s^*} \vartheta(\bx,s,t) \ ds \right)}_{= - \dt \nu - \dt \eta} + g(s^*,c) \vartheta(\bx,s^*,t) - g(s_0,c) \underbrace{\vartheta(\bx,s_0,t)}_{=h\nu} \\ + \underbrace{\nabla \cdot \left(\bu  \int_{s_0}^{s^*} \vartheta(\bx,s,t)   \ ds \right)}_{=\nabla \cdot \bu(1-\nu-\eta)}  = -\int_{s_0}^{s^*} \mu_v(s,c) \vartheta(\bx,s,t) \ ds.
\end{multline}
Substituting for $- \dt \nu - \dt \eta$ yields
\begin{equation}
\nabla \cdot \bu = b(c)\nu.
\end{equation}
Substituting $\bu= -\lambda \nabla p$ gives an equation for the pressure in $B_t$,
\begin{equation} \label{pressure}
-\lambda \nabla^2 p = b(c) \nu.
\end{equation}
Distributing the divergence operator gives
\begin{subequations} \label{forceBalanceFlux}
\begin{align}
\nabla \cdot (\bu \nu) = -\lambda \nabla p \cdot \nabla \nu + b(c) \nu^2, \\
\nabla \cdot (\bu \vartheta) = -\lambda \nabla p \cdot \nabla \vartheta+ b(c) \nu \vartheta, \\
\nabla \cdot (\bu \eta) = -\lambda \nabla p \cdot \nabla \eta+ b(c) \nu \eta.
\end{align}
\end{subequations}

We see from (\ref{pressure}) that $p$ is proportional to $\lambda^{-1}$, so that $\lambda \nabla p$ is independent of $\lambda$.  Consequently $\nu$, $\vartheta$, and $\eta$ are independent of $\lambda$, allowing us to set $\lambda=1$.

We impose periodic and other boundary conditions, similar to what was done in \cite{multispeciesBiofilm}, to obtain the complete model.  The active cell volume fractions satisfy
\begin{subequations}\label{model}
\begin{multline}
\dt \nu(\bx,t) - \nabla p \cdot \nabla \nu =  -\mu_v(c) \nu(\bx,t) + g(s^*,c) \vartheta(\bx,s^*,t)  \\ - h(c)\nu(\bx,t) + b(c)\nu(\bx,t)\big(1-\nu(\bx,t)\big), 
\end{multline}
for $x\in B_t$, $t>0$ with conditions
\begin{align}
\frac{\partial \nu}{\partial z} =0,& \qquad \bx \in \Gamma_B, t \geq 0, \\
\nu(\bx,0) = \nu_0(\bx),& \qquad \bx \in B_t,
\end{align} 
where $z$ denotes the spatial variable orthogonal to the surface $\Gamma_B$, and $\nu_0$ is the initial active cell population.  The dormant cell volume fractions satisfy
\begin{multline}
\dt \vartheta(\bx,s,t) + \ds(g(s,c) \vartheta(\bx,s,t)) - \nabla p \cdot \nabla \vartheta  = \\ -\mu_v(s,c) \vartheta(\bx,s,t) - b(c) \nu(\bx,t) \vartheta(\bx,s,t), 
\end{multline}
for $x\in B_t$, $s>s_0$,$t>0$, with conditions
\begin{align}
g(s_0,c)\vartheta(\bx,s_0,t)  = h(c) \nu(\bx,t), &  \qquad \bx \in B_t, t > 0,\\
\frac{\partial \vartheta}{\partial z} =0,& \qquad \bx \in \Gamma_B, t \geq 0, s > s_0, \\
\vartheta(\bx,s,0) = 0,& \qquad \bx \in B_t, s \geq s_0.
\end{align}
The fully inert cell volume fractions, including necrotic cells, satisfy
\begin{multline}
\dt \eta(\bx,t) - \nabla p \cdot \nabla \eta = \\ \mu_u(c) \nu(\bx,t) + \int_{s_0}^{s^*} \mu_v(s,c) \vartheta(\bx,s,t) \ ds - b(c) \nu(\bx,t) \eta(\bx,t),
\end{multline}
for $x\in B_t$, $t>0$, with conditions
\begin{align}
\frac{\partial \eta}{\partial z} =0,& \qquad \bx \in \Gamma_B, t \geq 0, \\
\eta(\bx,0) = \eta_0(\bx),& \qquad \bx \in B_t,
\end{align} 
where $\eta_0$ is the initial inert cell population.  Pressure satisfies
\begin{align}
-\nabla^2 p  = b(c) \nu, \qquad& \bx \in B_t, t\geq 0, \\
p = 0, \qquad& \bx \in \Gamma_t, t\geq 0, \\
\frac{\partial p}{\partial z} = 0, \qquad& \bx \in \Gamma_B, t\geq 0.
\end{align}
Let $f=[f_1,\ldots,f_m]$ and $c=[c_1,\ldots,c_m]$.   The chemical species satisfy, for $j=1,\ldots,m$,
\begin{align}
-D_j \nabla^2 c_j = f_j, \qquad& \bx \in \Omega, t>0, \\
 f_j = 0, \qquad& \bx \in \Omega\backslash B_t,\\
c_j =  c_j^*, \qquad& \bx \in \Gamma_{H_b}, t\geq 0, \\
\frac{\partial c_j}{\partial z} = 0, \qquad& \bx \in \Gamma_B, t\geq 0,
\end{align}
where the $D_j$ are chemical diffusion coefficients and the $c_j^*$ are the chemical concentrations in the bulk fluid.  The normal velocity of the interface $\Gamma_B$ is given by
\begin{equation}\label{fullDnormal}
-\nabla p \cdot {\bf n} = -\frac{\partial p}{\partial n},
\end{equation}
where ${\bf n}$ is the unit outward normal of $\Gamma_B$.

\end{subequations}

\section{Computations}\label{sec:computations}
In this section we present computational results for models of batch cultures, chemostat cultures, and biofilms.

\subsection{Batch Culture Dormancy As a Response to Nutrient Depravation}\label{sec:batch}
%Let $c= [c_1,c_2]^T$, where $c_1$ is oxygen concentration and $c_2$ is concentration of anti-microbial.   Define the total populations
%\begin{subequations}
%\begin{align}
%P_1(t) = \int_{s_0}^{s^*} \omega_1(s) v(s,t) \ ds, \\
%P_2(t) = \int_{s_0}^{s^*} \omega_2(s) v(s,t) \ ds, 
%\end{align}
%\end{subequations}
%where $\omega_1$ and $\omega_2$ represent the uptake ability of chemical species $c_1$ and $c_2$, resp., for cells with dormancy $s$.

\begin{figure}[t]
\centering
\subfigure[Total Cell Populations]{
\includegraphics[height=2.25in]{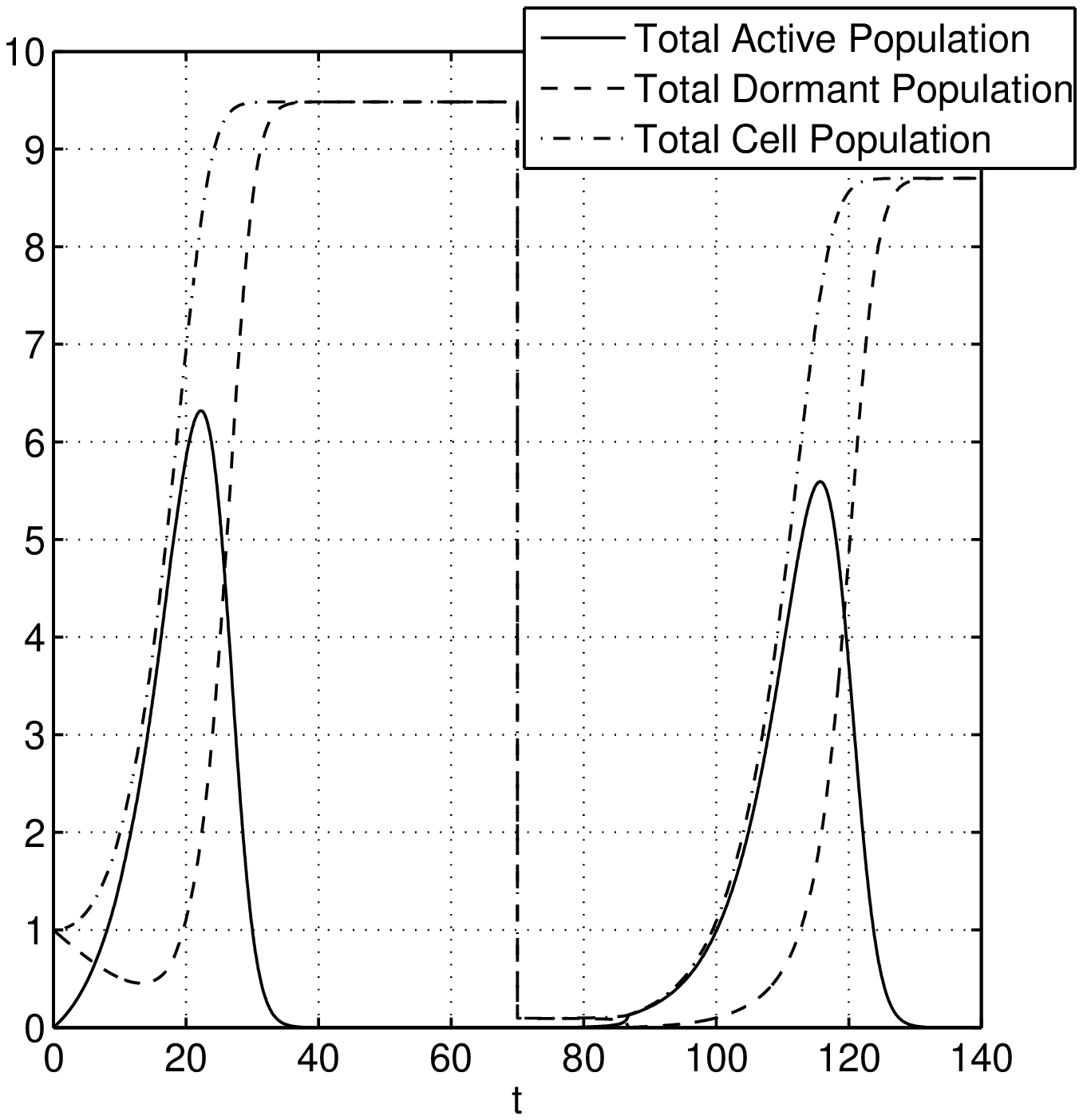}
\label{fig:cells}
}
\subfigure[Nutrient]{
\includegraphics[height=2.25in]{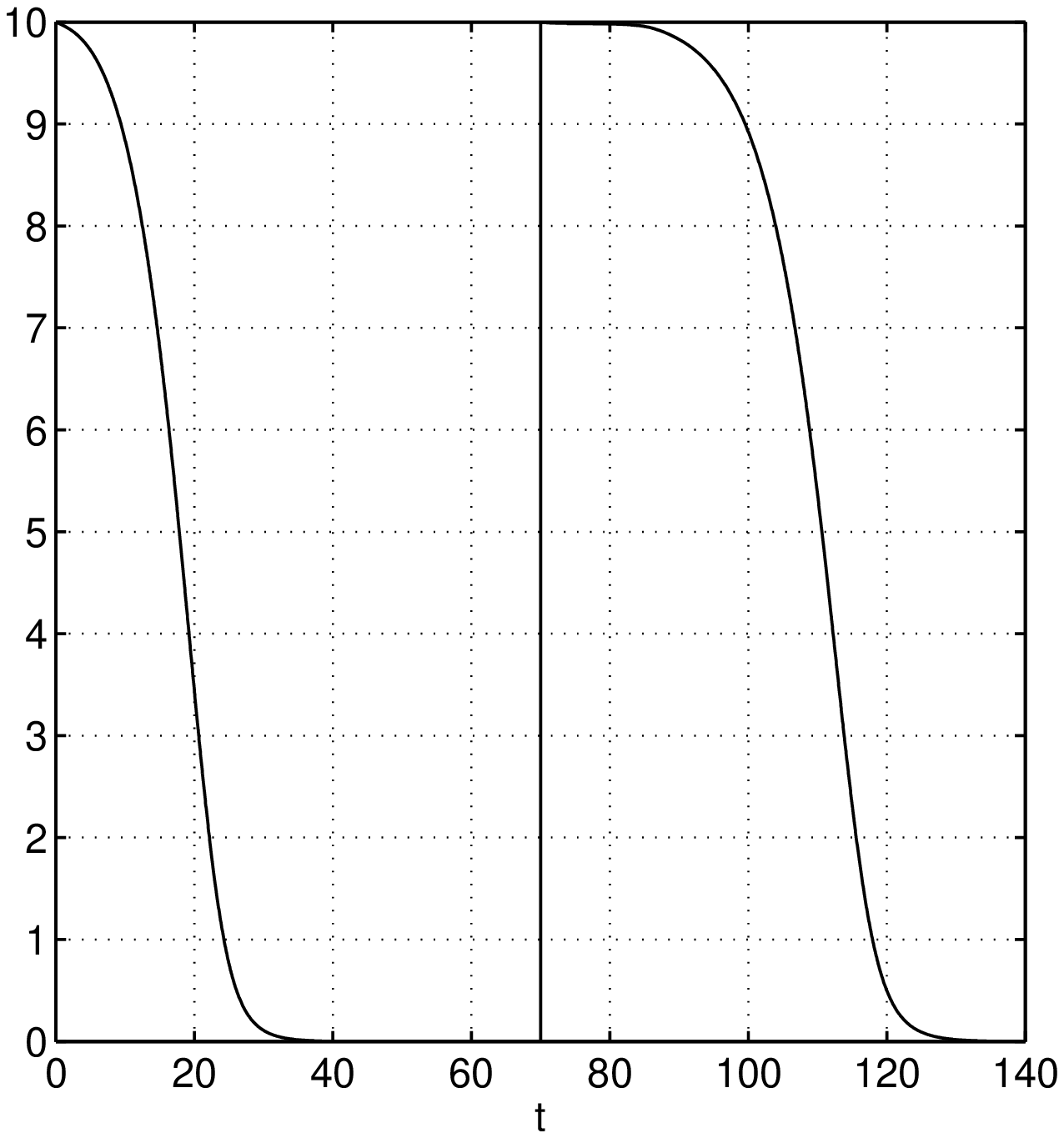}
\label{fig:nutrient}
}
\caption{Results for the batch model,  system (\ref{planktonic}) with  $d_0=0$.   We reculture $1\%$ of each subpopulation into new media at $t=70$ hours. Functional forms are as in (\ref{funcForms}) with parameters $k=1/4 \text{hr}$, $\gamma = 4 \text{g}_{\text{CODB}}/\text{m}^3$, $\epsilon_h = 0.05$, $\mu_u=0.005/\text{hr}$, $Y=0.63  \text{g}_{\text{CODB}}/\text{g}_{\text{CODS}}$, $\zeta = \text{g}_{\text{CODB}}/ 5 \text{m}^3$, $k_h = \text{g}_{\text{CODB}}/(6 \text{hr m}^3)$, and $k_g = 1/12$ hr.  The dormancy domain is $[0,1]$.  Time is measured in hours.}
\label{fig:batch}
\end{figure}

We let $c(t)$ be a scalar value representing nutrient.  We choose the functional forms
\begin{subequations}\label{funcForms}
\begin{align}
b(c) &= \frac{k c}{\gamma + c}, \\
h(c) &= \frac{k_h}{\zeta+c} + \epsilon_h, \\
g(s,c) &= \frac{k_g c}{\gamma + c}, \\
\mu_u(c) &= \mu_u,\\
\mu_v(s,c) &= 0,\\
f(c,u,v) &= -  \frac{k c}{Y(\gamma + c)}\left( u + \int_0^1 \big(e^{-s/k_g} + e^{-(1-s)/k_g} \big) v \ ds \right), \label{batchNutrient}\\
v_0(s) &= 0 \label{v_0},
\end{align}
\end{subequations}
with constant $\mu_u$, rate constants $k$, $k_h$, $k_g$, saturation constant $\gamma$, Monod constant $\zeta$, and yield constant $Y$.   We take the baseline parameters $k=1/4 \text{hr}$, $\gamma = 4 \text{g}_{\text{CODB}}/\text{m}^3$, $\epsilon_h = 0.05$, $\mu_u=0.005/\text{hr}$, and $Y=0.63  \text{g}_{\text{CODB}}/\text{g}_{\text{CODS}}$ \cite{Wanner06}.  The units $\text{g}_{\text{CODB}}$ and $\text{g}_{\text{CODS}}$ are the chemical oxygen demand of biomass and substrate mass, resp.    We assume a small $\zeta$, say $\zeta = \gamma/20 = \text{g}_{\text{CODB}}/ 5 \text{m}^3$.  We require  $k_h/\zeta \geq 1/24\text{hr}$, so take $k_h = \text{g}_{\text{CODB}}/(6 \text{hr m}^3)$.  We set the dormancy domain to be $s_0 = 0$ and $s^*=1$, and assume 24 hr emergence at nutrient saturation so that $k_g = 1/12$ hr.  As we reculture or restore nutrients, we see a reawakening of the population and a growth spurt until the new nutrient is also depleted.  Results are shown in Figure \ref{fig:batch} for reculturing of 1\% of each subpopulation onto new substrate at $t=70$ hours.

\begin{figure}[t]
\centering
\subfigure[Total Cell Populations]{
\includegraphics[height=2.25in]{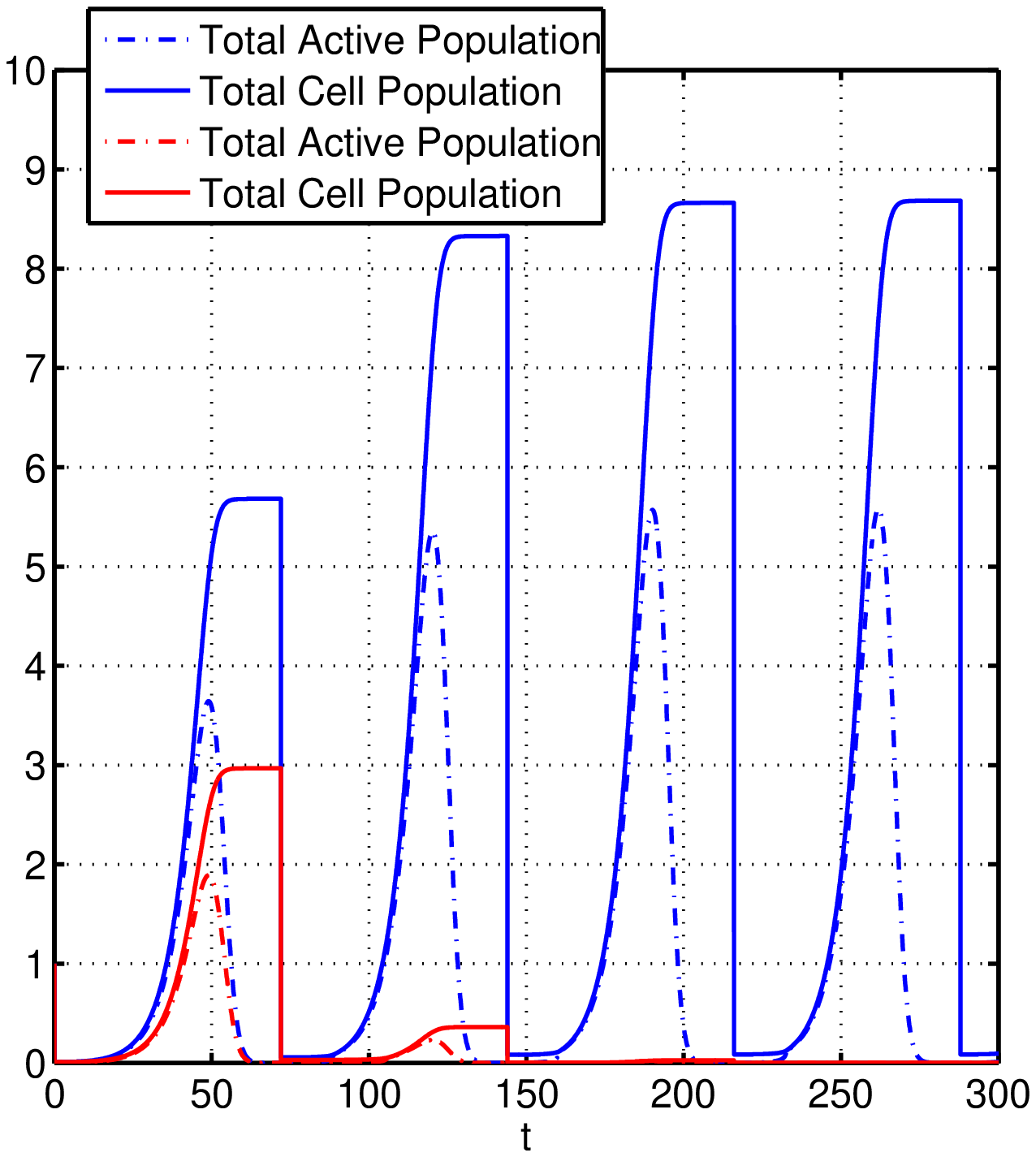}
\label{fig:batch72cells}
}
\subfigure[Nutrient]{
\includegraphics[height=2.25in]{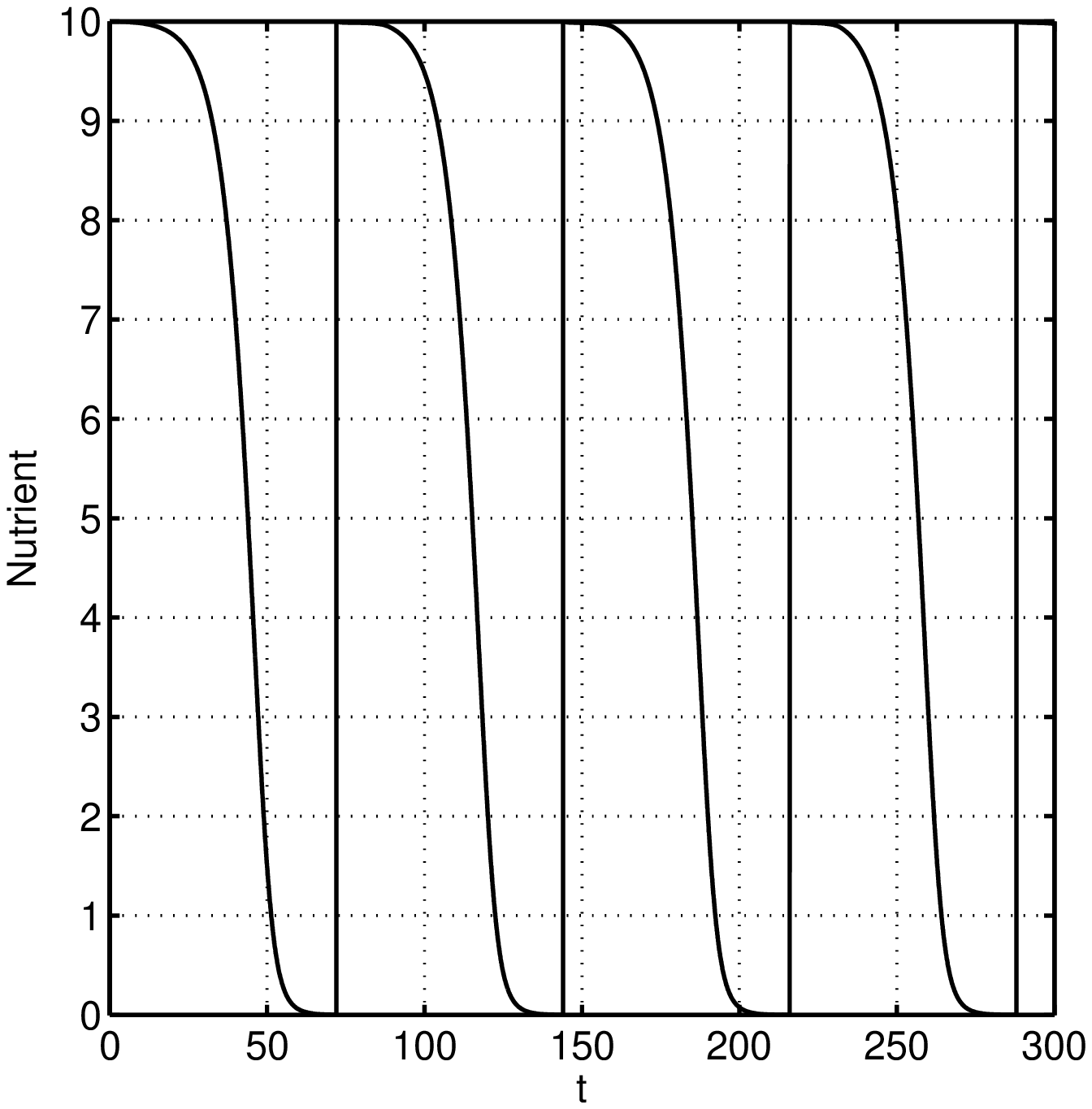}
\label{fig:batch72nutrient}
}
\caption{Results for the batch model for two species with $k_g=1/12$ (blue) and $k_g=1/24$ (red).   We reculture $1\%$ of each subpopulation into new media every $72$ hours.  All other parameters are as in Figure \ref{fig:batch}.}
\label{fig:batch_double}
\end{figure}

We conducted simulations with two species where $1\%$ of each subpopulation was recultured into fresh media every 48, 72, and 168 hours.   For shorter times of 4, 8, 16, and 24 (cases where $1\%$ reculturing leads to extinction of all species), we used $100\%$ reculturing.   In all cases the fast-waker population ($k_g=1/12$ hr) outgrew the slow-waker population ($k_g=1/24$ hr).  Although both active populations undergo oscillations, the fast-waker active population outgrows the slow-waker active population in each case.  Moreover, the fast-waker total population dramatically outgrows the slow-waker total population in each case.  (Results for 72-hour reculturing are shown in Figure \ref{fig:batch_double}.) This outcome is not surprising; conditions favor microbes that rapidly resuscitate.

%f(c,u,v) &=  \left[ 
%                           \begin{array}{cc} 
%                              - \frac{k_1 c_1u}{\gamma_1 + c_1} -  \frac{k_2 c_1 P_1}{\gamma_2 + c_1} \\
%                              - \frac{k_3 c_2u}{\gamma_3 + c_2} -  \frac{k_4 c_2 P_2}{\gamma_4 + c_2} 
%                            \end{array} 
%                      \right],  \\
%C_0(t) &=  \left[ 
%                           \begin{array}{cc} 
%                              C_o(t) \\
%                              C_p(t)
%                            \end{array} 
%                      \right],  \\
%{\bf I didn't do anything except removing the singularity from h(c) (that
%would definitely cause problems at least in the biofilm case). We''ll also want a time-dependence form for supplied oxygen (incidentally, oxygen
%may end up being carbon source or something) and antimicrobial, that is,
%for the upper boundary conditions on c in the biofilm case, and for the
%inflow concentrations in the chemostat case. To start, I would
%go with an on-off pulse for both, each with its own, adjustable frequency
%(frequency for the substrate may end up being 0). We may want to make
%the pulsing random at some point as well, likely with a Poisson distribution.}

\subsection{Chemostat Culture Dormancy As a Response to Nutrient Depravation}\label{sec:chemostat}

\begin{figure}[t]
\centering
\subfigure[Total Cell Populations]{
\includegraphics[height=2.25in]{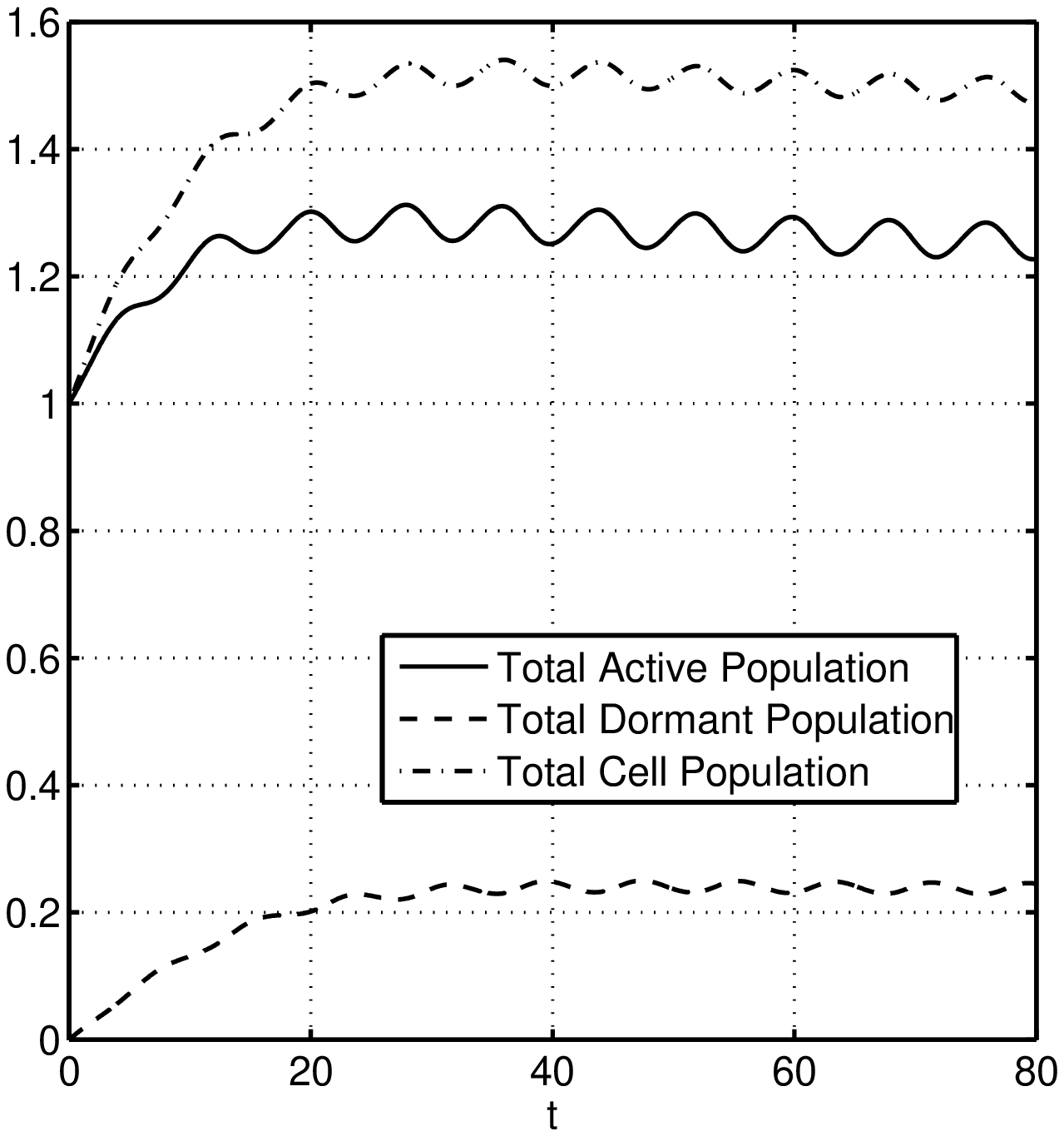}
\label{fig:cells_chemo}
}
\subfigure[Nutrient]{
\includegraphics[height=2.25in]{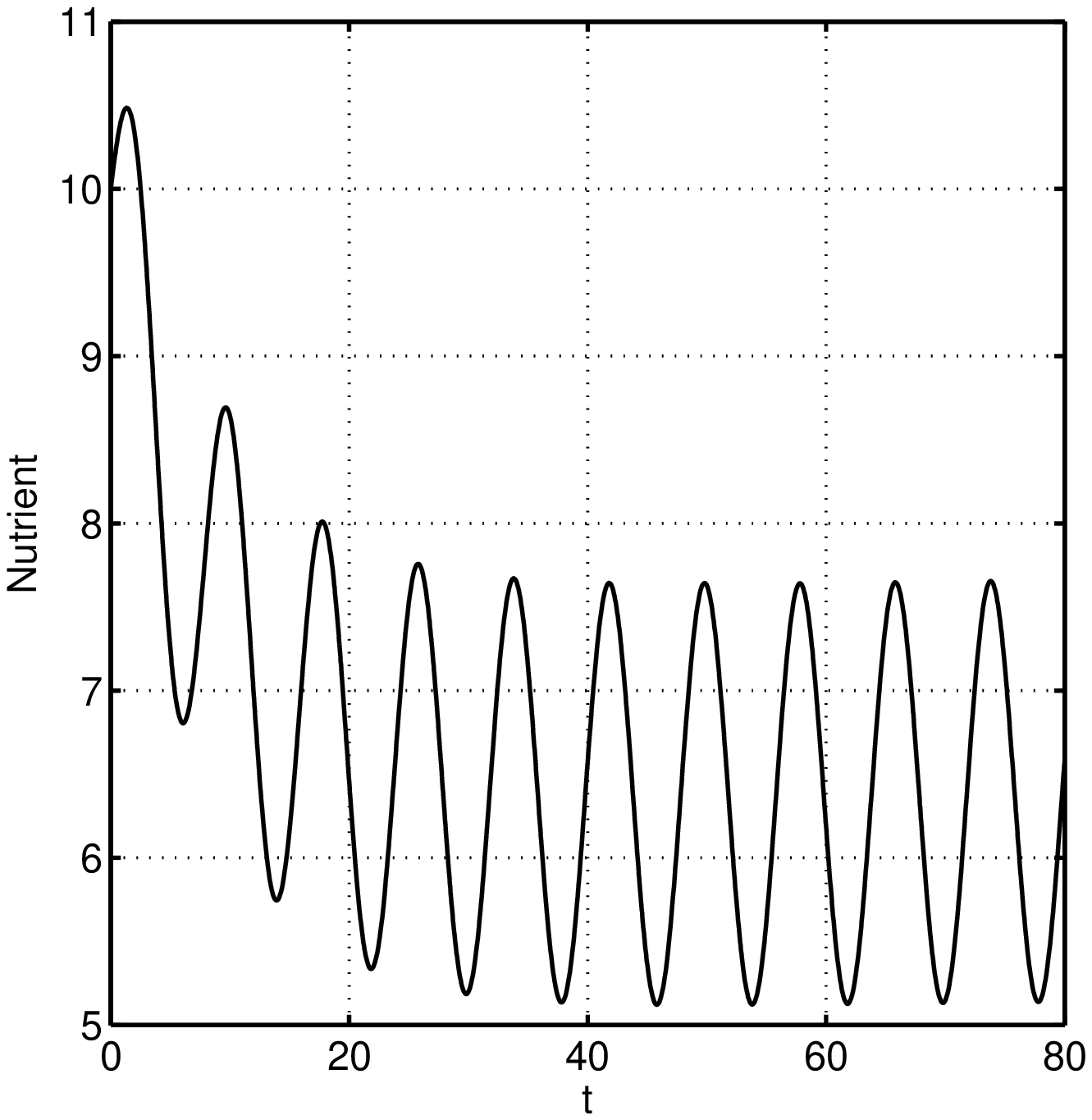}
\label{fig:nutrient_chemo}
}
\caption{Results for the chemostat model,  system (\ref{planktonic}) with $d_0=k/2$ and $C(t) = 8+8\cos(\pi t/4)$ where $t$ is measured in hours.   Functional forms are as in (\ref{funcForms}) with parameters $k=1/4 \text{hr}$, $\gamma = 4 \text{g}_{\text{CODB}}/\text{m}^3$, $\epsilon_h = 0.05$, $\mu_u=0.005/\text{hr}$, $Y=0.63  \text{g}_{\text{CODB}}/\text{g}_{\text{CODS}}$, $\zeta = \text{g}_{\text{CODB}}/ 5 \text{m}^3$, $k_h = \text{g}_{\text{CODB}}/(6 \text{hr m}^3)$, and $k_g = 1/12$ hr.  The dormancy domain is $[0,1]$.  Time is measured in hours.}
\label{fig:chemostat}
\end{figure}

We use the functional forms and parameter values of Section \ref{sec:batch}, with $d_0=k/2$ and $C_0(t) = 8 + 8 \cos(\pi t/4)$.  The active population, dormant population, and nutrient relax to a periodic oscillation.  Results are shown in Figure \ref{fig:chemostat}. 

\begin{figure}[t]
\centering
\subfigure[Total Cell Populations]{
\includegraphics[height=2.25in]{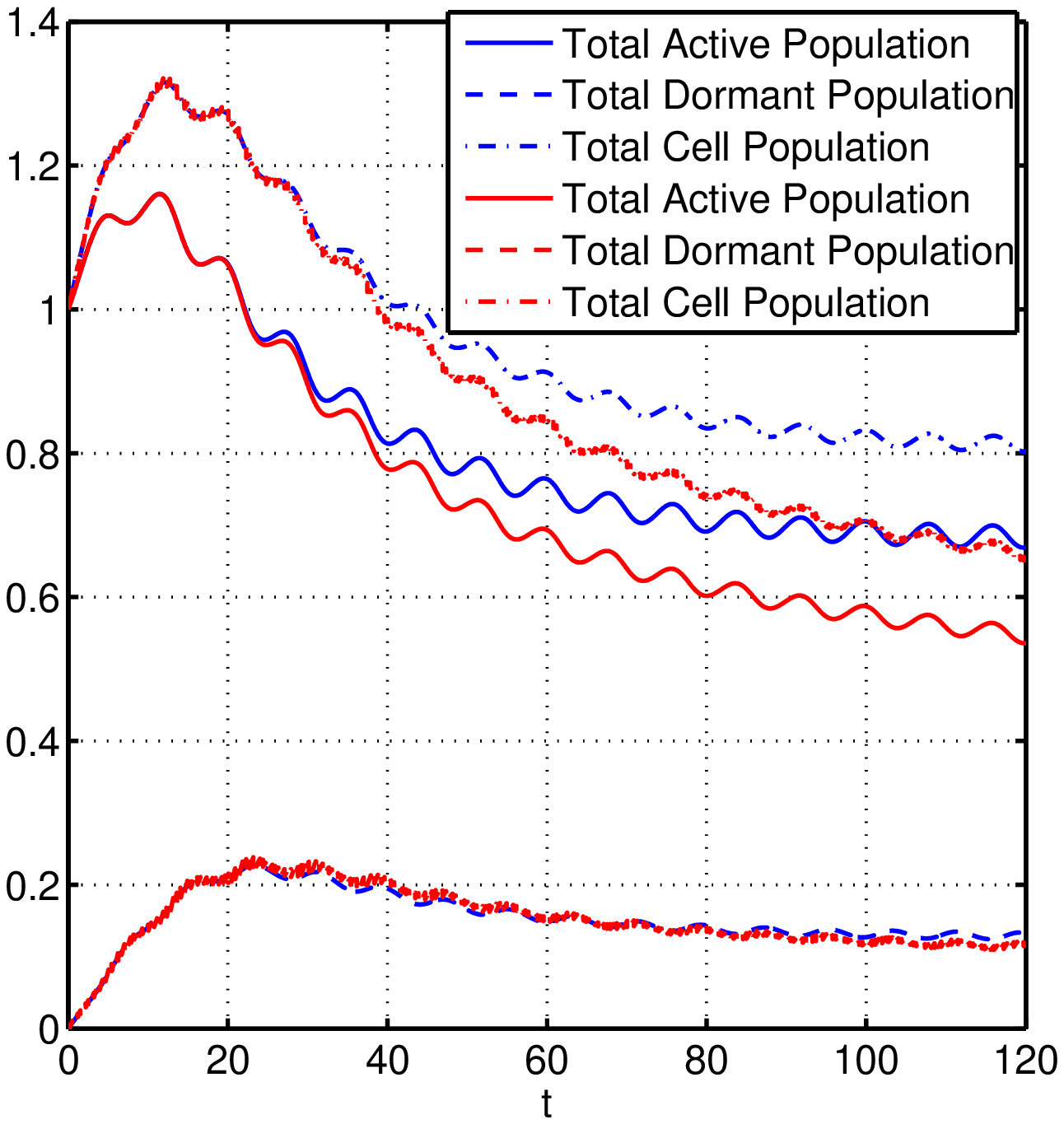}
\label{fig:cells_double}
}
\subfigure[Nutrient]{
\includegraphics[height=2.25in]{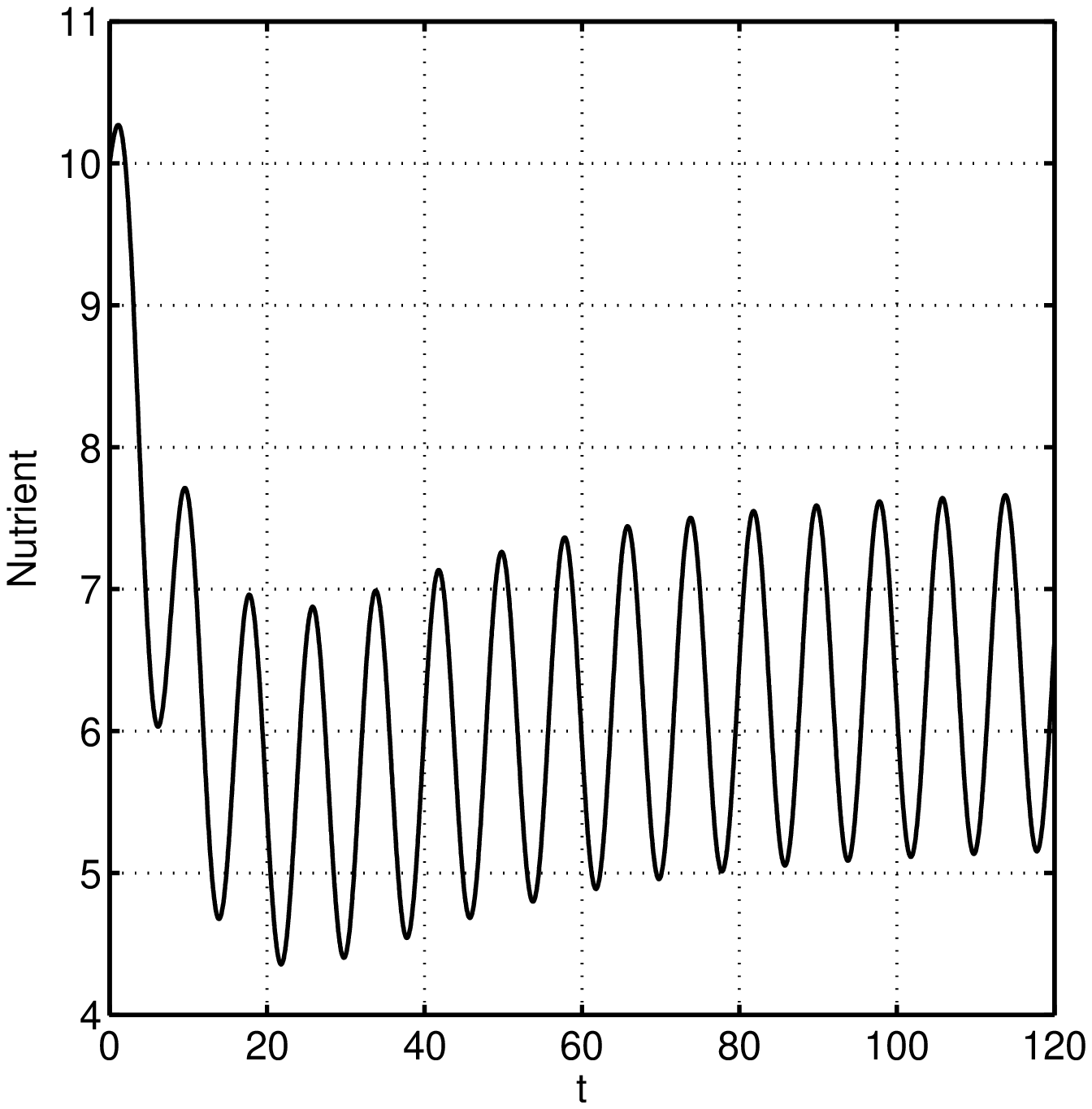}
\label{fig:nutrient_double}
}
\caption{Results for the chemostat model for two species with $k_g=1/12$ (blue) and $k_g=1/24$ (red).   All other parameters are as in Figure \ref{fig:chemostat}.  In the long run, the slow waker is driven to extinction.}
\label{fig:double}
\end{figure}

For two competing species, the only difference being $k_g=1/12$ vs.~$k_g=1/24$, the results are shown in Figure \ref{fig:double}.  As in the batch case, the faster waker outcompetes the slower waker in the long run.   This is true for a wide range of periods, $C(t) = 8+8\cos(2\pi t p)$ with $p$=0.5,1, 4, 12, 24, 48, 72 and 168 hours, verifying asymptotics predictions for short and long
periods and extending to intermediate periods.  Changes in period do not alter the numbers for a given subpopulation appreciably in magnitude, but rather change how they oscillate around some trajectory. The fast waker, as predicted, also outcompetes the slow waker in the case of a steady, rather than oscillating, nutrient source.

Computations of competing species in a chemostat, where one species undergoes no dormancy ($h=0$) and the other undergoes dormancy with parameter $k_g=1/12$, confirm the results of Section \ref{asymptoticsFast} (using $C(t) = 8+8\cos(4\pi t)$ where $t$ is measured in hours) and Section \ref{asymptoticsSlow} (using $C(t) = 8+8\cos(\pi t/4)$ where $t$ is measured in hours).    A species without dormancy capability will outcompete an otherwise similar species which can go dormant, under both fast and slow oscillations in nutrient.

\subsection{Biofilm Dormancy as a Response to Nutrient Depravation}\label{sec:biofilm}

In the biofilm model, (\ref{batchNutrient}) becomes
\begin{equation}\label{biofilmNutrient}
f(c,\nu,\vartheta) = -  \frac{k \rho^* c}{Y(\gamma + c)}\left( \nu + \int_0^1 \big(e^{-s/k_g} + e^{-(1-s)/k_g} \big) \vartheta \ ds \right).
\end{equation}
We use the functional forms and parameter values of Section \ref{sec:batch}, with the addition of $\rho^*= 10^4 \text{g}_{\text{CODB}}/\text{m}^3$ and $D=10^{-4} \text{m}^2/\text{day}$ \cite{Wanner06}.

\begin{figure}[t]
\centering
\subfigure[$k_g = 1/12$ hr]{
\includegraphics[height=2.25in]{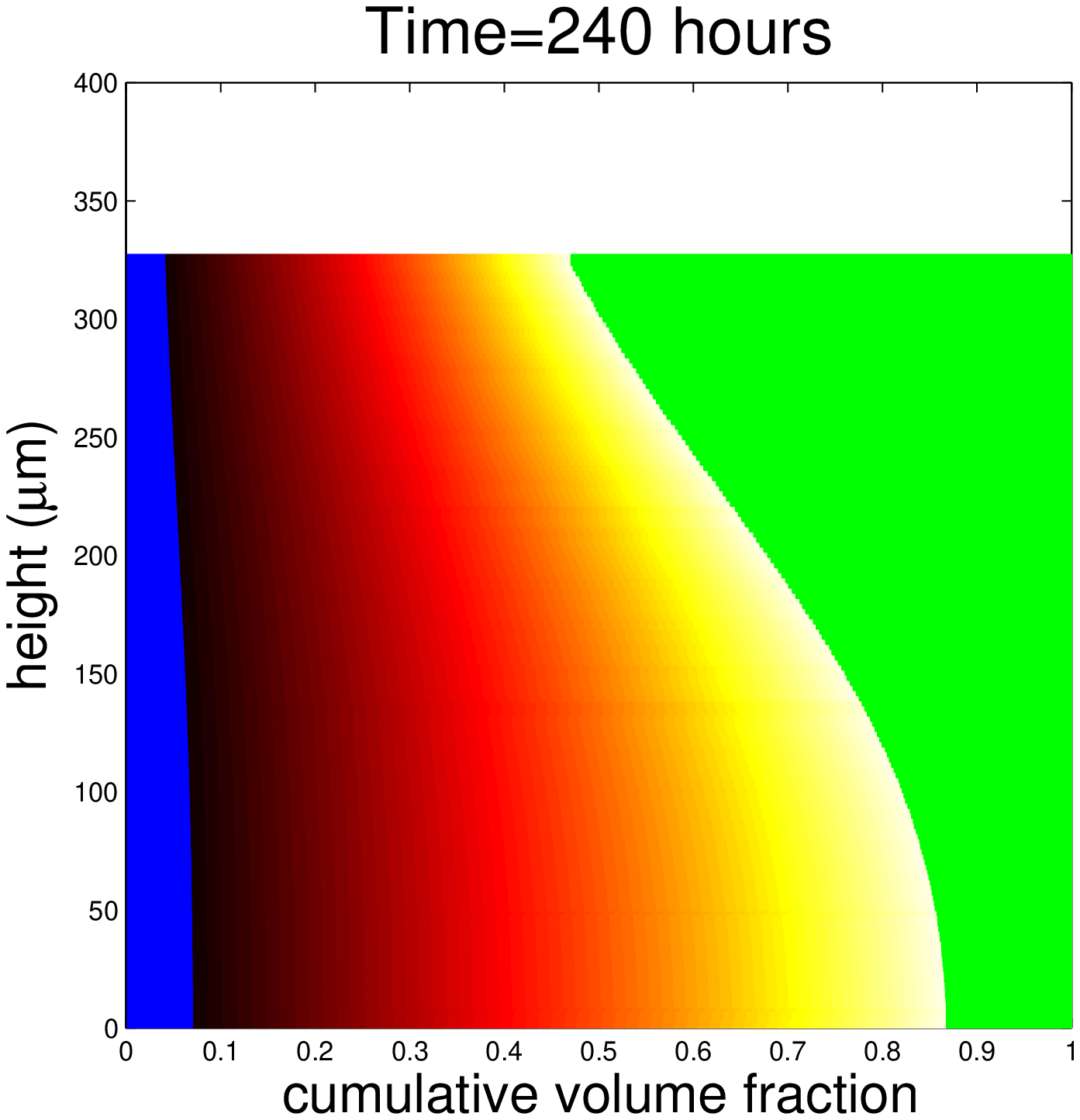}
\label{fig:biofilm12}
}
\subfigure[$k_g = 1/24$ hr]{
\includegraphics[height=2.25in]{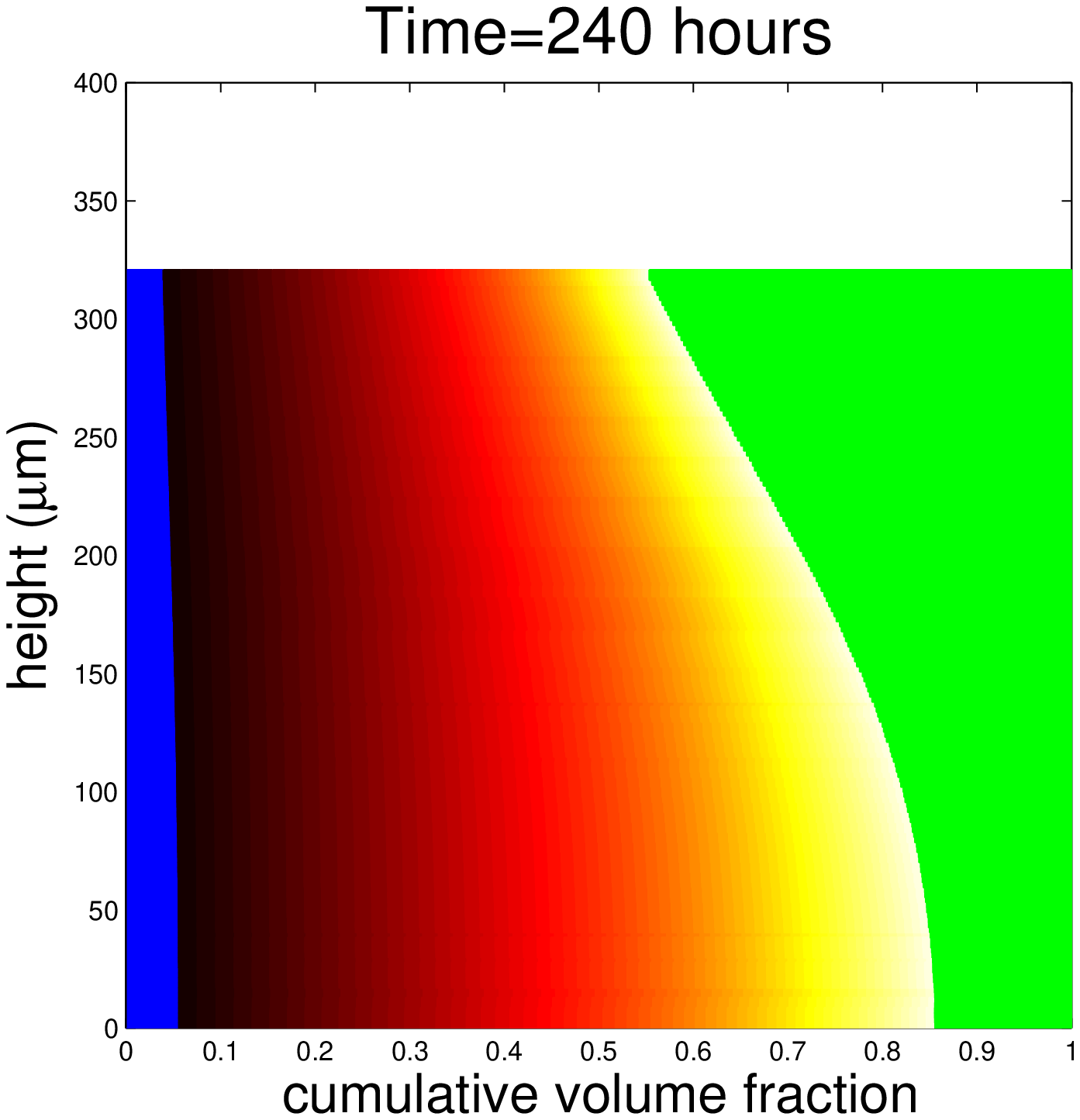}
\label{fig:biofilm24}
}
\caption{Results for the biofilm model,  system (\ref{model}).     Functional forms are as in (\ref{funcForms}) except that $f$ given by (\ref{biofilmNutrient}).  Parameters $k=1/4 \text{hr}$, $\gamma = 4 \text{g}_{\text{CODB}}/\text{m}^3$, $\epsilon_h = 0.05$, $\mu_u=0.005/\text{hr}$, $Y=0.63  \text{g}_{\text{CODB}}/\text{g}_{\text{CODS}}$, $\zeta = \text{g}_{\text{CODB}}/ 5 \text{m}^3$, $k_h = \text{g}_{\text{CODB}}/(6 \text{hr m}^3)$.  The two figures differ in the values of $k_g$. The dormancy domain is $[0,1]$.  Time is measured in hours. The horizontal width of a color constitutes the
volume fraction of cells of in the corresponding state.  Green denotes active cells, blue denotes fully inert cells, and the ``hot" black-red-yellow-white spectrum denotes dormancy from 0 to 1.  While fast reactivators are more prevalent near the top of the biofilm, slow reactivators have more biomass, particularly in dormant (hence resistant) form, in deeper biofilm layers.}
\label{fig:biofilm}
\end{figure}

Results shown in Figure \ref{fig:biofilm} indicate that the slow waker has comparable total live biomass than the fast waker, and possibly more in lower regions, and palpably more dormant biomass, even though the faster wakers outnumber the slower wakers near the top of the biofilm.

\section{Computational Methodology}

As was done for senescence-structure in \cite{MMSbiofilm07,persistenceSenescence}, we often handle a general physiologically structured system such as~(\ref{sPlanktonic})-(\ref{sBirth})
more easily if it is transformed to an age-structured
system, whether in the statement of the problem, or indirectly in the numerical method \cite{deRoos97}.  Let $a \geq 0$ represent time a cell has spent dormant.  We make a change of variables so that dormancy, $s(a,t)$, is a separate function of
age and time. We then get age-structured equations for $v$,
\begin{subequations}
\begin{align}
\dt v(a,t) + \da v(a,t) = -\tilde{\mu}_v(s,c) v(a,t) - Dv(a,t),& \quad 0 < a \leq a^*, t>0,\\
v(0,t) = h(c) u(t), & \quad t > 0, \\
\dt s(a,t) + \da s(a,t) = g(s(a,t),c(t)), & \quad 0 < a \leq a^*,  t>a,\\
s(0,t) = s_0,&
\end{align}
\end{subequations}
where $\mt_v(s,c) = \mu_v(s,c) + \ds g(s,c)$ and $s(a^*,t) = s^*$.   The condition (\ref{v_0}) means we need only consider $t>a$ for the domain of $s(a,t)$.

For our choice, $g(s,c) = \frac{k_g c}{\gamma_g + c}$ for scalar $c$, we get 
\begin{equation}
s^* =  s_0 + \int_{t-a}^t \frac{\gamma_g + c(\tau)}{k_g c(\tau)} s^* \ d\tau,
\end{equation}
so that $a^* \rightarrow \infty$ if $c \rightarrow 0$.   Since functions with similar behavior to $g$ are natural representations of the dormancy dynamics, we find that the original physiologically structured system is more tractable computationally than the equivalent age-structured system for most forms of $g$ that interest us.

To solve equations with more general physiological structure, we use an extension of the natural-age-grid Galerkin methods developed for age- and space-structured systems in \cite{age-pwconst-paper,age-general-paper}.  These methods move the discretization nodes in age smoothly along characteristic lines.  The solutions are approximated by piecewise polynomials, rather than moments as was done by de Roos \cite{deRoos89}.   Our extension of our methods to general physiological structure moves the discretization nodes in the physiological variable along characteristic curves, similar to a method of Sulsky \cite{sulsky-mass}, but with the preservation of the property in our methods that each time step need not result in a new discretization node in the physiological variable.  This is essential when variation of spatial structure, or any other dynamics in the problem, occurs on a faster time scale than that of the physiological trait.   Otherwise, the need to take lots of small time steps would induce many more physiological nodes than are necessary for accuracy, resulting in potentially great loss of efficiency from additional computation or interpolation onto a coarser grid. 

To motivate the integration in age and time, we ignore for the moment the discretization in space.  We partition the domain $[s_0,s^*]$ at each time by the set of nodes $\left\{s_i(t)\right\}_{i=0}^{N}$ where $s_0(t) = s_0$.  If $s_N(t) \geq s^*$, we simply ignore that node and the function value over it until needed.  This is not an issue for our choices of $g$.    We compute the solution at times ${t_j}$ and let $\Delta t_j = t_{j+1}-t_j$,  $s_{i,j}=s_i(t_j)$, and $\Delta s_{i,j} = s_{i+1,j} - s_{i,j}$.   For the last interval we use $\Delta s_{N,j} = \max(s^*-s_{N,j},0)$.  Although we are not including space in this discussion for reasons of clarity, the presence of spatial structure in a problem will induce different time scales into a problem, making adaptivity and nonuniformity of time intervals an important property of any method used.

For the computations in this paper, we use a piecewise constant approximation space over the domain $[s_0,s^*]$.  Higher-order approximation spaces can be used, as was done in age in \cite{age-pwconst-paper}.  We define the projection into the space of piecewise constants over the partition of $[s_0,s^*]$ by
\begin{equation}
\Pi(v(s,t_j)) = \left\{
                           \begin{array}{rl}
                           \frac{1}{\Delta s_{i,j}} \int_{s_{i,j}}^{s_{i+1,j}} v(s,t_j) \ ds,& \qquad \text{if } s_{i,j} \leq s < s_{i+1,j}, \\
                           0,& \qquad \text{otherwise}.
                           \end{array}
	              \right.
\end{equation}

We make the approximation $V_{i,j} \approx \Pi(v(s,t_j))$ via the following algorithm.   Let $\Delta s_{\text{max}}$ be the largest we want the first interval in $s$ to be. For most time steps we have $\Delta s_{0,j} \leq \Delta s_{\text{max}}$.  In this case we set
\begin{subequations}
\begin{align}
s_{i,j+1} = s_{i,j} + \Delta t_j g(s_{i,j},c(t_j)), \qquad  \text{for } i=1,\ldots,N.
\end{align}
We choose $\Delta t_j$ such that $s_{i,j+1} \geq s^*$ for at most one $i$, so as to keep $N$ fixed.

Let the value $V_{i, j}$ denote the density over $[s_{i,j},s_{i+1,j}]$ for $i=1,\ldots,N-1$.   We use ${\mathcal B(t)} $ to denote the creation of newly dormant cells at $s_0$.  Then
\begin{align}
V_{i,j+1} = \frac{\Delta s_{i,j}}{\Delta s_{i,j+1}} V_{i,j},& \qquad \text{for } i=1,\ldots,N-1, \\
V_{0,j+1} = \frac{1}{\Delta s_{0,j+1}}\big( \Delta s_{0,j} V_{0,j} + \Delta t_j{\mathcal B(t_j)}\big),& \\
V_{N,j+1} = \frac{\Delta s_{N-1,j} - \Delta t_j}{\Delta s_{N,j+1}} V_{N-1,j}.&
\end{align}
Because the applications in this paper provide for the first extension of the methods presented in \cite{age-pwconst-paper,age-general-paper}, we have kept ${\mathcal B(t)}$ general in this part of the presentation of the method.    In our case we have  ${\mathcal B(t_j)}=g(s_0,c(t))v(s_0,t)$.  Also, if $g$ is independent of $s$, we have $\Delta s_{i,j}/\Delta s_{i,j+1} = 1$ for $i=1,\ldots,N-1$.

If $\Delta s_{0,j} > \Delta s_{\text{max}}$, we introduce a new node and set
\begin{align}
s_{i+1,j+1} = s_{i,j} + \Delta t_j g(s_{i,j},c(t_j)),& \qquad \text{for } i=1,\ldots,N-1,\\
V_{i+1,j+1} = \frac{\Delta s_{i,j}}{\Delta s_{i,j+1}} V_{i,j},& \qquad \text{for } i=0,\ldots,N-1,
\end{align}
for the intermediate intervals, and set
\begin{align}
s_{1,j+1} &= \Delta t_j g(s_0,c(t_j)), \\
V_{0,j+1} &= \frac{\Delta t_j {\mathcal B(t)}}{\Delta s_{0,j+1}},\\
V_{N,j+1} &= \frac{ \Delta s_{N-1,j} V_{N-1,j} + \big( \Delta s_{N,j}  -  \Delta t_j g(s^*,c(t_j)) \big) V_{N,j}}{\Delta s_{N,j+1}},
\end{align}
\end{subequations}
for the first and last intervals. 

The above calculations account for transport in the physiological variable, entry into dormancy, and exit from dormancy.    Upwind differences approximate the advection terms in space.  Center differences approximate the diffusion terms in space.  Backward Euler formul{\ae}, embedded in step-doubling with local extrapolation, approximate the time derivatives.  This creates a likely second-order correct time integration scheme \cite{step-doubling-paper}.

\section{Conclusions}

Modeling results suggest that spatial heterogeneity in biofilms can support a rich dormancy structure.  For example, whereas dormant cells near the top of a biofilm would need to be able to resuscitate quickly (small $s^*$, large $g$) when environmental conditions improve in order to be competitive, dormant cells lower in the biofilm, where the slower waker has a defensive advantage over the fast waker due to a larger amount of dormant biomass without an appreciable difference in total live biomass, may be able to afford to be more cautious (large $s^*$, small $g$).

In contrast, dormancy-capable cells in well-mixed, planktonic systems (e.g. batch and chemostat cultures) appear to have less advantage over ``regular'' cells.  In the absence of spatially structured populations, live biomass is maximized by the fastest possible exit from dormancy.    The lower limit of time to reawakening is governed by physiological, biochemical or other constraints within the cells, and hence dormancy mechanisms are constrained to easily reversible mechanisms. As most lab populations are of the well-mixed batch or chemostat sort, and most natural populations are of the spatially-structured biofilm sort, this presents a possible drawback in use of typical laboratory systems for characterization of natural ones.

We remark that we have only considered here dormancy response in the context
of resource depravation. Dormancy is also likely an effective defense
strategy against antimicrobial agents -- many antimicrobials are only
effective against metabolically active targets. Thus the presence of
antimicrobials reinforces the utility of dormancy in biofilms and
also may advantage dormancy-capable populations in well-mixed cultures.
The nature of dormancy as defense could itself benefit from modeling
studies.

More generally, beyond dormancy specifically,
recent studies suggest that phenotypic heterogeneity of
many sorts is typical in spatially structured microbial populations
such as biofilms \cite{BolesPNAS2004}. Hence, methods of the sort presented
here are likely to be useful and possibly necessary for modeling the
function and ecology of spatially unmixed microbial populations of the sort
that dominate the natural environment.

\section*{Acknowledgments}  The authors thank Phil Stewart for helpful
ideas and discussions. 

\bibliographystyle{plain}
\bibliography{all}

\end{document}